\documentclass[12pt]{iopart}

\usepackage{graphicx}

\begin{document}
\title{Dynamics of a single vortex line in a condensate}
\author{V. Bretin, P. Rosenbusch, and J. Dalibard}
\address{Laboratoire Kastler Brossel$^*$, 24 rue Lhomond,
75005 Paris, France}

\begin{abstract}
{We study experimentally the line of a single vortex in a rotating
prolate Bose-Einstein condensate confined in a harmonic potential.
In agreement with predictions, we find that the vortex line is in
most cases curved at the ends. We also present measurements of the
quadrupole oscillation of the condensate in presence of a single
vortex. A theoretical treatment to account for the short time and
long time behaviors of these modes is developed. Finally, we use
these measurements to infer the angular momentum per particle and
relate it to the shape of the vortex line.}
\end{abstract}



\section{Introduction}

The rotation of a macroscopic quantum object is a source of
spectacular behaviors. A fascinating phenomenon is the nucleation
of vortices under rotation. In superfluid liquid helium contained
in a cylindrical bucket rotating around its axis $z$, one observes
the nucleation of quantized vortices for a sufficiently large
rotation frequency $\Omega$ \cite{Donnelly91}. A similar
phenomenon occurs in Bose-Einstein condensates confined in a
rotating harmonic trap
\cite{Cornellphaseimprinting,Madison00,Ketterle,Hodby01,Cornellcooling}.
These quantized vortices in superfluids are analogous to the
quantized flux lines in superconductors~\cite{supra}. Among the
remaining problems, is the shape of a vortex/flux line and the
study of its time evolution. The observation of inclined flux
lines in superconductors was only possible due to recent advances
in electron microscopy \cite{fluxlineMFM}. In gaseous condensates,
one has comparably easy access to the vortex line because the
density of the atom cloud is low. A few disordered vortex lines
have been observed by taking tomographic images perpendicular to
the long axis of a cigar shaped condensate \cite{Ketterle}. An
array of more than 30 vortex lines in a pancake shaped condensate
has been observed by transverse imaging of the whole atom cloud
\cite{Engels02}.

In this article we report the full length observation of a single
vortex line in a cigar shaped condensate. We find that as a result
of spontaneous symmetry breaking the line is generally
bent~\cite{Rosenb02}. Our experimental results confirm recent
predictions, in which the shape of the vortex line minimizing the
energy of the gas was derived for a given rotation frequency
\cite{Garcia01a,Garcia01b,Aftalion01,Modugno,Aftalion02}.

We also present measurements of the angular momentum of the
rotating condensate, based upon surface wave spectroscopy. We
develop the theoretical framework to account for the excitation of
the quadrupole surface modes of the condensate. We then relate the
measured angular momentum of the gas to the bending of the vortex
line and its deviation from the center.

\section{Preparation of a single vortex}

\subsection{Experimental setup}

Our $^{87}$Rb condensate is formed by radio-frequency (rf)
evaporation of $10^9$ magnetically trapped atoms in the $F=m_F=2$
state. The longitudinal and transverse frequencies of the trap are
respectively $\omega_z/2\pi=11.8$~Hz and $(\omega_x+\omega_y)/4\pi
= \omega_\bot/2\pi \sim 100$~Hz (the $x$ axis is vertical).
Because gravity slightly displaces the center of the trap with
respect to the magnetic field minimum, the potential in the $xy$
plane is not perfectly isotropic and we measure a 1\% relative
difference between $\omega_x$ and $\omega_y$.

The condensation threshold is reached at $T_{\rm c}\sim300$~nK,
with $N_{\rm c}\sim2\times 10^6$~atoms. We cool to typically
$T\sim 90$~nK which corresponds to a condensate with
$N_0\sim5\times 10^5$~atoms and a chemical potential $\mu\sim
70$~nK. It is obtained using $\nu_{\rm f}=\nu_0 +10$~kHz as the
final rf, where $\nu_0$ is the frequency at which the trap is
emptied. During the rest of the experimental cycle, we maintain
the evaporation rf at an adjustable level, typically $\nu=\nu_0
+12$~kHz, to control the temperature.

\subsection{Vortex nucleation}

Once the condensate is formed, we use an off-resonant laser beam
to impose on the trapping potential an elliptic anisotropy in the
$xy$ plane \cite{Madison00}. The wavelength of the beam is 852~nm,
its power 0.1~mW and its waist $20\;\mu$m. Acousto-optic
modulators deflect the position of the beam in the $xy$ plane,
thereby rotating the potential anisotropy at a frequency of
$\Omega/2\pi=70$~Hz. The potential created by the stirrer can be
written~:
\begin{equation}
U({\bf r})=\frac{\epsilon}{2}M\omega_\perp^2
\;\left(X^2-Y^2\right)\ , \label{eq:pot}
 \end{equation}
where the axes $X,Y$ are deduced from the fixed $x,y$ axes by a
rotation by an angle $\Omega t$. $M$ is the atomic mass and the
coefficient $\epsilon$ measures the relative strength of the
stirring and the magnetic potential. We use typically $\epsilon
\sim 4\%$.

We apply this ``laser stirrer" for 300~ms, during which $\sim$ 7
vortices are nucleated \cite{Madison01}. The condensate then
evolves freely in the magnetic trap for an adjustable time $\tau$.
The atom distribution at $\tau$ is probed destructively by
switching off the magnetic trap, letting the cloud expand during
$t_{\rm TOF}=25$~ms and performing absorption imaging.

\begin{figure}
\centerline{\includegraphics[width=15cm]{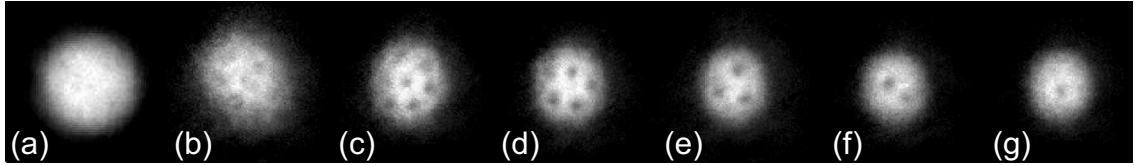}}
 \caption{Image of the
condensate with the probe beam propagating along the $z$ axis. a)
Before stirring. b) Just after stirring ($\tau=0$). c)
$\tau=150$~ms. d) $\tau=500$~ms. e) $\tau=800$~ms. f)
$\tau=1200$~ms. g) $\tau=1800$~ms. } \label{fig:latticedecay}
\end{figure}

The preparation of a single vortex line takes advantage of the
slight static anisotropy of our magnetic trap, so that the angular
momentum is not exactly a constant of motion. In a time $\tau \sim
1$--$2$~s, we observe a transition from a multi-vortex condensate
to a condensate with a single vortex (see
Fig.~\ref{fig:latticedecay}). This relatively long time levels the
fluctuations that may occur during the nucleation process. Thereby
we are able to reproduce a condensate with a single vortex on
every experimental cycle. This vortex line can then be studied for
a time $\tau \leq 10$~s.

\begin{figure}[b]
 \centerline{\includegraphics[height=5cm]{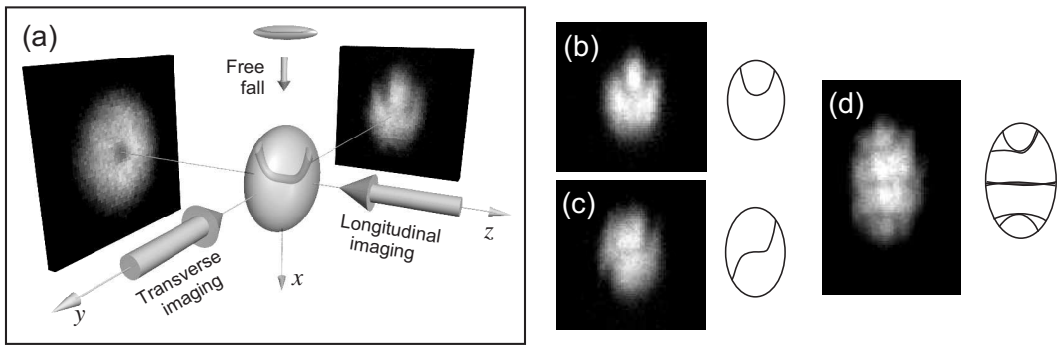}}
\caption{(a) Schematic of the imaging system allowing for an
observation of the full vortex line. The initially cigar shaped
condensate is imaged after 25~ms of time of flight leading to the
inversion of the ellipticity in the $xz$ plane. Two beams image
the atom cloud simultaneously along the longitudinal ($z$) and
transverse ($y$) directions of the initial cigar. (b), (c)
Transverse images of a condensate with a single vortex line
respectively with a ``U" and ``N" shape. The schematic represents
the form of the vortex line. (d) Transverse image of a condensate
with 7 vortices.} \label{fig:images}
\end{figure}

In order to obtain an image of the full vortex line, we send two
imaging beams aligned along the $y$ and $z$ directions, which
probe the atom cloud simultaneously (Fig.~\ref{fig:images}a). The
beams are combined onto a camera with the same magnification.
During the expansion, the transverse dimensions $x$ and $y$ of the
condensate are magnified by $\omega_\bot t_{\rm TOF}\sim 15$,
while the longitudinal dimension is nearly unchanged
\cite{Castin96}. It has been shown theoretically that the presence
of a single vortex line does not alter this expansion and that the
coordinates of the line are scaled by the same factors
\cite{Modugno}.

The transverse images in Fig.~\ref{fig:images}b,c respectively
obtained for $\tau=7.5$~s and $\tau=5$~s show the vortex line as a
line of lower atomic density. Clearly this vortex line is not
straight. It rather has the shape of a wide ``U" or an unfolded
``N". The fraction of N vs. U shaped vortices depends on
temperature. Mostly U's are observed at the lowest temperature
($T=75$~nK), while $\sim$ half of the vortices are N-shaped for
$T\geq 100$~nK. Fig.~\ref{fig:images}d has been obtained with a
7-vortex lattice (one vortex at the center and the six other ones
at the summits of an hexagon, such as in
Fig.~\ref{fig:latticedecay}c). We have selected an image for which
the lattice planes are parallel with the transverse imaging beam.
The dark trace passing through the center of the condensate is the
sum of 3 vortex lines. At the top and bottom, 2 lines lie on top
of each other.

\begin{figure}
\scalebox{0.995}{\centerline{\includegraphics[height=4cm]{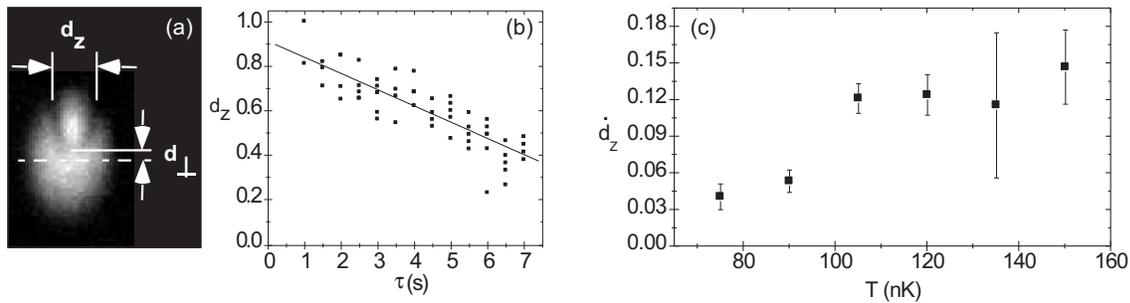}}}
 \vskip -0.3cm
\caption{(a) Schematic of the extraction of $d_z$ and $d_\bot$
from the vortex line. (b) Evolution of $d_z$ with $\tau$. Each
point corresponds to a single image. (c) Variation of $\dot d_z$
with the temperature $T$.}
 \label{fig:reproducibility}
\end{figure}

\subsection{Evolution of a single vortex}

In order to give a quantitative analysis of the time evolution of
a condensate with a single vortex, we measure the distance along
the $z$ direction between the two points where the vortex line
leaves the condensate (Fig.~\ref{fig:reproducibility}a). Note that
the ends of the vortex line in the transverse image are nearly
always well visible with good contrast ($\sim 30\,\%$), even if
the contrast of the line at the center of the condensate is poor
(below $20\,\%$). Therefore this measurement can be performed on
most images. Normalization by the length of the condensate along
$z$ leads to the quantity $d_z$.

In Fig.~\ref{fig:reproducibility}b, we plot $d_z$ as a function of
$\tau$. Each point in the plot corresponds to a single image. The
normalized length of the vortex line $d_z$ decreases
quasi-linearly with time. The graph shows that after $\tau=7$~s
the vortex is still present (in 7~s the number of atoms is divided
by a factor 2.5). This long lifetime is reminiscent of the MIT
result \cite{Ketterle}, where a vortex array with more than 100
vortices was produced at $\tau=0$. The number of vortices was
divided by $4$ in $\sim 5$~s; however a single vortex could still
be detected after $\tau=40$~s. In both experiments, it is clear
that the decay time of the last vortex is much longer than the one
of the initial array.

We have repeated this experiment for different temperatures $T$
during the free evolution time $\tau$.
Fig.~\ref{fig:reproducibility}c displays the measured slopes as a
function of temperature. We observe a quasi-linear decrease of
$d_z$ for all temperatures. Our results can be compared to the one
reported in \cite{Aboshaeer02}, where the decay rate of an array
of $\sim\,100$ vortices as a function of temperature was measured.
In \cite{Aboshaeer02}, a 60\% increase in temperature leads to an
increase by a factor 17 in the decay rate. This variation is more
dramatic than ours, indicating different decay mechanisms for a
single vortex and for a large vortex array. In the latter case,
the friction between the vortices and the thermal cloud makes the
two components of the system stick together \cite{Fedichev01}. In
the case of a single vortex, the most probable scenario is that
the rotation of the thermal component rapidly stops due the
residual static trap anisotropy \cite{dgo}. The vortex line is
then dragged to the edge of the condensate due to the friction
between the two components \cite{Fedichev}.

\section{Angular momentum of a condensate with a single vortex}

An important parameter for the characterization of a rotating
condensate is the average angular momentum per particle $ L_z$.
This quantity is related to the transverse quadrupole frequencies
of the condensate $\omega_{\pm }$ by the formula~:
\begin{equation}
L_z=\frac{1}{2}\; M r_\perp^2\; \left(\omega_{+} -
\omega_{-}\right) \ , \label{lift}
\end{equation}
where $r_\perp^2$ stands for the average value of $x^2+y^2$ for
the condensate. The physical basis of (\ref{lift}) is clear. In
absence of angular momentum along the $z$ axis, the system is
rotationally invariant and the two transverse quadrupole modes
$m=+2$ and $m=-2$ have the same frequencies $\omega_+=\omega_-$
(and also the same damping rates $\gamma_+=\gamma_-$). In presence
of a vortex the symmetry between the modes $m=\pm 2$ is broken
 and (\ref{lift}) relates the corresponding lift of degeneracy
$\omega_+-\omega_-$ to $L_z$.

The result (\ref{lift}) was established in \cite{Stringari} for a
rotating condensate with an arbitrary number of vortices in the
Thomas Fermi limit. For a condensate with a well centered vortex
it had been obtained in \cite{Dodd97,Sinha97,Svidzinsky98}. In
this section we discuss the measurement of $\omega_{\pm }$ based
on a percussional excitation of the condensate and we present the
relation between the results of our measurements of $L_z$ and the
shape of the vortex line.

\subsection{Measurement of $\omega_{\pm }$ from the
precession of the condensate axes}

We first prepare the condensate with a single vortex as described
above. We then excite it by applying the laser potential with
fixed axes ($X,Y=x,y$ in Eq. (\ref{eq:pot})) for a short duration
$t_0=0.5$~ms $\ll \omega_\bot^{-1}$. Thereby, the wave function
$\psi_0$ of the condensate is transformed into
 \begin{equation}
  \psi({\bf r})=\psi_0({\bf r})+ \delta \psi({\bf r})
  \qquad \mbox{with} \qquad
  \delta \psi({\bf r})=-\frac{i t_0}{2\hbar}
  \epsilon M \omega_\bot^2(x^2-y^2)\ .
 \end{equation}
This excites a superposition of $m=+2$ and $m=-2$ modes with equal
amplitudes. More precisely, $\delta \psi$ can be written in terms
of the usual $u_\pm$ and $v_\pm$ functions characterizing the
quadrupole modes $m=\pm 2$ in a Bogoliubov analysis (see e.g.
\cite{Ohberg})~:
\begin{equation}
\delta \psi({\bf r}) =i\epsilon' \left(u_+({\bf r}) +v_+^*({\bf
r})\right) + i\epsilon' \left(u_-({\bf r})+v_-^*({\bf r})\right)\
,
\end{equation}
where $\epsilon'$ is a dimensionless coefficient proportional to
the strength $\epsilon$ and to the duration $t_0$ of the
excitation.

We then let the cloud evolve freely in the magnetic trap for a
variable time $t$. The condensate wave function becomes:
\begin{eqnarray}
\psi({\bf r},t)=e^{-i\mu t/\hbar}\Big\{ \psi_0({\bf r}) &+& i
\epsilon' e^{-\gamma_+ t}\left(e^{-i\omega_+ t}u_+({\bf
r})+e^{i\omega_+ t}v_+^*({\bf r})  \right) \nonumber \\
&+& i \epsilon' e^{-\gamma_- t}\left(e^{-i\omega_- t}u_-({\bf
r})+e^{i\omega_- t}v_-^*({\bf r})  \right) \Big\}\ ,
\label{eq:psilong}
\end{eqnarray}
where $\mu$ is the chemical potential of the condensate. In
(\ref{eq:psilong}) we added a phenomenological damping term for
each mode. We then perform the time of flight analysis and measure
the size $R_{\rm l}$ and the polar angle $\theta$ of the long
axis, the size $R_{\rm s}$ of the short axis, and calculate the
cloud ellipticity $\zeta=R_{\rm l}/R_{\rm s}$. After some
manipulation, we find that the predicted values for these
quantities can be written in the following form:
\begin{equation}
\cos 2\theta =\frac{F(t)}{\sqrt{F^2(t)+G^2(t)}}\qquad \sin 2\theta
=\frac{G(t)}{\sqrt{F^2(t)+G^2(t)}} \label{theta}
\end{equation}
and
\begin{equation}
\zeta-1 \propto \sqrt{F^2(t)+G^2(t)}\ , \label{zeta}
\end{equation}
where the functions $F$ and $G$ are defined by
\begin{eqnarray}
F(t)&=&e^{-\gamma_-t}\sin(\omega_-t)+e^{-\gamma_+t}\sin(\omega_+t)\
,
\\
G(t)&=&e^{-\gamma_-t}\cos(\omega_-t)-e^{-\gamma_+t}\cos(\omega_+t)
\ .
\end{eqnarray}

\subsection{Various time domains for the condensate precession}

We now discuss the  behavior of the precessing condensate expected
from Eqs. (\ref{theta}-\ref{zeta}), assuming that the damping
rates $\gamma_\pm$ of the quadrupole modes are much smaller than
the frequencies $\omega_\pm$.

\subsubsection{Short time behavior.}

For times short compared to $\gamma_\pm^{-1}$, the exponential
decay of the functions $F$ and $G$ can be neglected and we obtain
simply:
\begin{eqnarray}
&& \theta = \frac{\left(\omega_+-\omega_-\right)t}{4}\qquad
\mbox{modulo }\frac{\pi}{2}\ , \label{theta2} \\
&&\zeta-1 \propto \left| \sin \left( (\omega_++\omega_-)t/2\right)
\right|\ .
\end{eqnarray}
In this case, the lift of degeneracy $\omega_+-\omega_-$ is simply
obtained from the measurement of the precession angle $\theta$.
The oscillation of the ellipticity is a function of the average
frequency $(\omega_++\omega_-)/2$.

We note that combining (\ref{lift}) and (\ref{theta2}), we obtain
the result
\begin{equation}
\theta=\frac{L_z t}{2M r_\perp^2}\ . \label{theta0}
\end{equation}
For very short times (i.e. $\omega_\pm t \ll 1$) this result is
actually quite general, as we show below. It applies to any fluid
confined in a harmonic potential with binary interactions,
provided the fluid is initially in a stationary state with
rotational invariance, described by a density operator $\rho_0$:
\begin{eqnarray*}
&&\langle x^2 \rangle_0 =\langle y^2 \rangle_0=r_\perp^2/2 \qquad
\langle xy \rangle_0=0 \qquad \langle p_x p_y\rangle_0=0\\
&&\langle x p_x +p_x x \rangle_0=\langle y p_y +p_y y
\rangle_0=0\qquad \langle xp_y +y p_x \rangle_0=0
\end{eqnarray*}
Note that the system can have some angular momentum ($L_z=\langle
x p_y -p_y x \rangle_0\neq 0$). After the percussional excitation
by the laser potential varying as $x^2-y^2$, the new density
operator of the system is at first order in the perturbation:
\begin{equation}
 \rho=\rho_0 - i\frac{\eta}{2}
\left[ \sum_{i=1}^{N} x_i^2 -y_i^2, \rho_0 \right]\ , \label{rho}
\end{equation}
where the sum runs over the $N$ particles of the fluid. $x_i$ and
$y_i$ denote the position operators of the $i$-th particle in the
$xy$ plane. The real coefficient $\eta$ is a measure of the
strength of the percussional excitation. We now calculate the time
evolution of $\langle x^2 \rangle_t$, $\langle y^2 \rangle_t$ and
$\langle xy \rangle_t$. Using Ehrenfest theorem we obtain:
\[
\frac{d}{dt}\langle x^2 \rangle=\frac{i}{N\hbar}\sum_{i=1}^N
\langle [H,x_i^2]\rangle = \frac{1}{m}\langle p_x x + x p_x
\rangle\ .
\]
The average on the right hand side is calculated in the state
(\ref{rho}) and we get for short times~:
\begin{equation}
\langle x^2\rangle_t =\frac{r_\perp^2}{2}-\frac{\hbar \eta
t}{M}r_\perp^2\qquad ,\qquad \langle y^2\rangle_t
=\frac{r_\perp^2}{2}+\frac{\hbar \eta t}{M}r_\perp^2\ .
\label{gene1}
\end{equation}
The calculation of $\langle xy\rangle$ is slightly more involved.
It requires the value of $\langle x p_y+yp_x\rangle$ to first
order in time and in $\eta$. We use the fact that in equilibrium
$\langle xF_y+yF_x\rangle_0=0$, where ${\bf F}({\bf r})$ is the
force created on a given particle located in ${\bf r}$ by the
$N-1$ remaining particles. We then obtain
\begin{equation}
\langle xp_y+yp_x\rangle_t=-\frac{2\,\eta\hbar L_z \,t}{M} \qquad
, \qquad \langle x y\rangle_t =-\frac{\eta \hbar L_z \,t^2}{M^2}\
. \label{gene2}
\end{equation}
Quite remarkably the results (\ref{gene1}-\ref{gene2}) are
independent of the binary interaction potential between the
particles. The polar angle $\theta$ characterizing the long axis
of the condensate is obtained from
\begin{equation}
\tan 2\theta =\frac{2\langle x y\rangle_t}{\langle
x^2\rangle_t-\langle y^2\rangle_t}\ ,
\end{equation}
which is indeed equivalent to (\ref{theta0}). The present proof,
valid for any fluid confined in a harmonic potential with binary
interactions, holds only for $t\ll \omega_\perp^{-1}$. As stated
above, (\ref{theta2}) holds also for much longer times
($\omega_\perp^{-1} \leq t \ll \gamma_\pm^{-1}$) for a pure BEC in
the Thomas-Fermi limit \cite{Stringari}.

\subsubsection{Long time behavior.}

At longer times, the exponential decays of the amplitudes of the
modes cannot be neglected anymore. If the two quadrupole modes
decay with the same rate ($\gamma_+=\gamma_-$), the result
(\ref{theta2}) is valid at any time. Simply the amplitude $\zeta$
of the quadrupole oscillation becomes weaker and weaker, which
makes it harder to detect. On the contrary, if the two rates
$\gamma_+$ and $\gamma_-$ are different, say e.g. $\gamma_+ <
\gamma_-$, one reaches at long time a regime where the mode $m=+2$
has an amplitude is much larger than $m=-2$. In this case one
simply gets $\theta=\omega_+ t/2$ and $\zeta-1 \propto
e^{-\gamma_+ t}$.

\subsection{Experimental results}

\begin{figure}
 \centerline{\includegraphics[height=5cm]{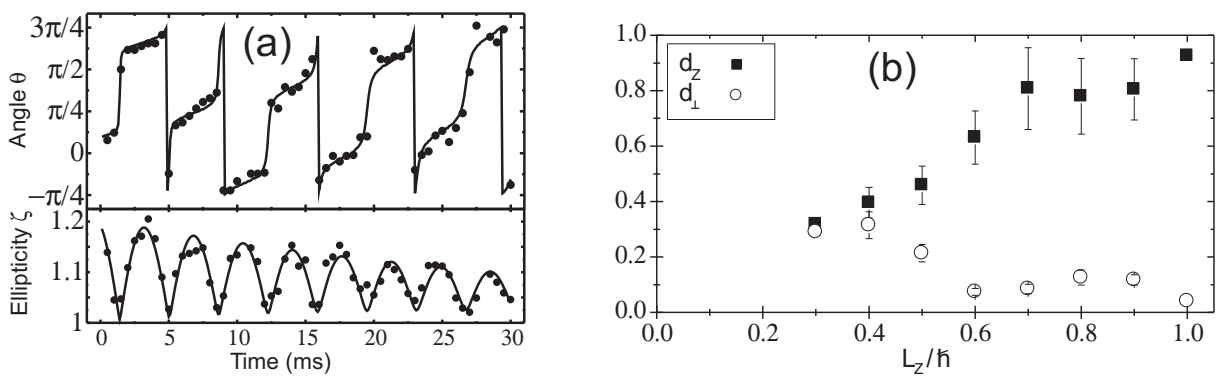}}
 \caption{(a) Angle $\theta$ and ellipticity $\zeta$ as a function
 of time $t$ in presence of a single vortex. The value of the angular
momentum per particle $L_z$ is deduced from the precession of the
main axes observed for small $t$. (b) Variation of $d_z$ and
$d_\bot$ as functions of $L_z$. All measurements were binned into
intervals of $L_z=0.1\;\hbar$ and averaged.}
 \label{film}
\end{figure}

Typical evolutions of the quantities $\theta$ and $\zeta$ are
plotted in Fig.~\ref{film}a. They have been obtained for a
transverse frequency $\omega_\perp/2\pi=98.5$~Hz and a nearly
straight vortex ($\tau=2$~s). We observe the two successive
regimes anticipated above. For short times, one observes the
quadrupole oscillation with precessing axes. For $t>5$\,ms, the
precession rate increases and the jumps of $\pi/2$ in $\theta$
become more and more rounded. This behavior indicates that the
mode $m=-2$ decays faster than $m=+2$, as discussed above. The fit
using (\ref{theta})-(\ref{zeta}) gives: $\omega_{+}/2\pi= 159.5\pm
1.0\,$Hz, $\omega_{-}/2\pi= 116.8\pm 1.0\,$Hz, $\gamma_{+}=19.1\pm
2.0\,$s$^{-1}$ and $\gamma_{-}=35.7\pm 4.0\,$s$^{-1}$. We note
that the two measured frequencies satisfy the sum rule
$\omega_{+}^2+\omega_{-}^2=4\,\omega_{\perp}^2$ \cite{Stringari}
with a good accuracy. From the difference $\omega_+-\omega_-$ and
the measured size of the condensate, we infer that the angular
momentum per particle is $L_z\sim \hbar$ (to within 10\%)
\cite{lz}. The study of the difference between $\gamma_+$ and
$\gamma_-$ is outside the scope of this paper and it will be
discussed elsewhere \cite{Bretin03}.

\subsection{Relationship between $L_z$ and the shape of the vortex
line}

If we are interested only in $L_z$, we do not have to repeat the
full experimental sequence leading to Fig.~\ref{film}a. It is
sufficient to perform only three experimental runs, which all
start by preparing a condensate with a single vortex as described
before. For the first two runs, we excite the two quadrupole
surface modes $m=\pm 2$ and we probe the quadrupole oscillation at
the two first maxima of the ellipticity $\zeta$ after the flash.
From the precession of the condensate axes, we infer the frequency
difference between the quadrupole modes, hence $L_z$
\cite{Stringari,lz}. For the third run, we repeat the experiment
without the laser flash and analyze the shape of the vortex line.
As above we extract the normalized length $d_z$ of the vortex line
along the $z$ axis. We also measure the displacement $d_\bot$ of
the axial part of the vortex line (bottom of the U) from the
center of the condensate, and normalize it by the radius of the
condensate in the $xy$ plane (Fig.~\ref{fig:reproducibility}a).
Since we have access only to the projection of the decentering on
the $xz$ plane, we actually measure $d_\bot|\cos \alpha|$, where
$\alpha$ is the azimuthal angle of the axial part of the line. We
account for this geometrical factor by dividing the measured
displacement by $\langle |\cos \alpha|\rangle=2/\pi$.

Fig.~\ref{film}b shows $d_z$ and $d_\bot$ as functions of
$L_z/\hbar$. For clarity we group all points into bins of
$L_z=0.1\;\hbar$ and average over $d_z$ and $d_\bot$. The error
bars give the statistical variation. The data corresponding to
$L_z\leq0.2\;\hbar$ are not reproducible enough and are omitted. A
straight ($d_z\sim 1$) and well centered ($d_\bot\ll 1$) vortex
line corresponds to $L_z\sim \hbar$. The graph shows that
$d_z\simeq L_z/\hbar$. When $L_z$ decreases, we measure a
decentering $d_\bot\leq 0.15$ as long as $L_z>0.5\;\hbar$. Below
$L_z=0.5\;\hbar$, $d_\bot$ rises to 0.3.

\subsection{Comparison with previous theoretical works}

We now compare our experimental results with recent predictions
for the shape of a vortex line in an inhomogeneous cigar shaped
condensate
\cite{Garcia01a,Garcia01b,Aftalion01,Modugno,Aftalion02}. These
theoretical studies consist in looking for the ground state of the
condensate in a frame rotating at an angular frequency $\Omega$.
The general conclusion is that above a critical frequency
$\Omega_{\rm c}$ the ground state of the system has one or several
vortices. The central vortex is generally bent if the trap aspect
ratio $\omega_\bot/\omega_z$ is large compared to 1, which is the
case in our experiment. A precursor of this bending effect can
also been found in \cite{Svidzinsky,Feder} in which a stability
analysis of a straight vortex in an elongated condensate showed
that some bending modes have negative energy, and are thus
unstable.

Our experimental procedure is somewhat different from the one
considered in these theoretical studies. $L_z$ and $\Omega$ play
the role of a couple of conjugate extensive/intensive variables
just like for instance volume and pressure. In our case no
rotating anisotropy is imposed onto the condensate during the
relevant evolution. The stirring laser has been switched on for a
short time only at the beginning of the procedure in order to set
a non-zero angular momentum in the system. We observe the
evolution of the condensate in our static trap, as the angular
momentum of the gas slowly decays. At first sight, neither $L_z$
seems to be conserved nor $\Omega$ imposed by the experimental
environment. However, the long lifetimes observed for the vortex
(up to 8 s) suggest that during an interval of a fraction of a
second (a typical thermalization time), $L_z$ is almost constant.
Therefore the description of our rotating condensate at time
$\tau$ should rather correspond to a system with a fixed $L_z$
than to a system rotating with a fixed $\Omega$. The states of
minimal energy may differ between the two descriptions
\cite{Modugno,Rokshar}. More precisely states with an angular
momentum $L_z$ notably below $\hbar$ are found as saddle points of
the energy functional calculated in the rotating frame (fixed
$\Omega$), while they appear as energy minimum when one considers
a condensate as ours, with a fixed angular momentum $L_z$ and
where the $\Omega$ parameter is in principle not relevant
\cite{Modugno}.

\section{Conclusion}

We have reported the complete observation of a single quantized
vortex line. The U-shaped vortex line that we observe is
remarkably similar to those predicted and plotted in
\cite{Garcia01a,Garcia01b,Aftalion01,Modugno}. As pointed out in
\cite{Garcia01b}, the bending of the vortex line is a symmetry
breaking effect which does not depend on the presence of a
rotating anisotropy and which happens even in a completely
symmetric setup. We have given a detailed theoretical account of
the relation between measured condensate deformations and the
quadrupole mode characteristics (frequency and damping rate). The
quadrupole oscillations of the rotating condensate were used to
measure the angular momentum. The shape of the vortex line
(bending and deviation from the center) were then related to the
angular momentum of the system. Our results should help modelling
the dissipative evolution of a rotating Bose-Einstein condensate.

We thank Y. Castin, F. Chevy, and G. Shlyapnikov for useful
discussions. P. R. acknowledges support the Alexander-von-Humboldt
Stiftung and by the EU (contract number HPMF CT 2000 00830). This
work was partially supported by CNRS, Coll\`{e}ge de France,
R\'egion Ile de France, DGA, DRED and EC (TMR network ERB
FMRX-CT96-0002).


\vskip 1cm

\begin{thebibliography}{99}
\bibitem[*]{byline}
Unit\'e de Recherche de l'Ecole normale sup\'erieure et de
l'Universit\'e Pierre et Marie Curie, associ\'ee au CNRS.



\bibitem{Donnelly91}
R.J. Donnelly, {\it Quantized Vortices in Helium II}, (Cambridge,
1991).


\bibitem{Cornellphaseimprinting}
M.R. Matthews {\it et al.}, Phys. Rev. Lett. {\bf 83}, 2498
(1999).


\bibitem{Madison00}
K.W. Madison {\it et al.}, Phys. Rev. Lett. {\bf 84}, 806, (2000).


\bibitem{Ketterle}
J.R. Abo-Shaeer \emph{et al.}, Science \textbf{292}, 476 (2001);
C. Raman {\it et al.}, Phys. Rev. Lett. {\bf 87}, 210402 (2001).


\bibitem{Hodby01}
E. Hodby \emph{et al.}, Phys. Rev. Lett. \textbf{86}, 2196 (2001).

\bibitem{Cornellcooling}
P.C. Haljan {\it et al.}, Phys. Rev. Lett. {\bf 87}, 210403
(2001).

\bibitem{supra}
M. Tinkham, \emph{Introduction to superconductivity} (McGraw-Hill,
1996).

\bibitem{fluxlineMFM}
A. Tonomura \emph{et al.}, Nature \textbf{412}, 620 (2001); A.
Tonomura \emph{et al.}, Phys. Rev. Lett. \textbf{88}, 237001
(2002).


\bibitem{Engels02}
P. Engels \emph{et al.}, Phys. Rev. Lett. \textbf{89}, 100403
(2002).

\bibitem{Rosenb02} P. Rosenbusch, V. Bretin, and J. Dalibard,
Phys. Rev. Lett. \textbf{89}, 200403 (2002).

\bibitem{Garcia01a}
J.J. Garc\'{\i}a-Ripoll and V.M. P\'erez-Garc\'{\i}a, Phys. Rev. A
\textbf{63}, 041603 (2001).


\bibitem{Garcia01b}
J.J. Garc\'{\i}a-Ripoll and V.M. P\'erez-Garc\'{\i}a, Phys. Rev. A
\textbf{64}, 053611 (2001).


\bibitem{Aftalion01}
A. Aftalion and T. Riviere, Phys. Rev. A \textbf{64}, 043611
(2001).


\bibitem{Modugno}
M. Modugno \emph{et al.}, Eur. Phys. J. D \textbf{22}, 235 (2003).


\bibitem{Aftalion02}
A. Aftalion and R. L. Jerrard, cond-mat/0204475.


\bibitem{Madison01}
K.W. Madison {\it et al.}, Phys. Rev. Lett. {\bf 86}, 4443 (2001).


\bibitem{Castin96}
Y. Castin and R. Dum, Phys. Rev. Lett. {\bf 77}, 5315 (1996).


\bibitem{Aboshaeer02}
J.R. Abo-Shaeer \emph{et al.}, Phys. Rev. Lett. \textbf{88},
070409 (2002).


\bibitem{Fedichev01}
O.N. Zhuravlev \emph{et al.},
Phys. Rev. A \textbf{64}, 053601 (2001).


\bibitem{dgo}
D. Gu\'ery-Odelin, Phys. Rev. A \textbf{62}, 033607 (2000).


\bibitem{Fedichev}
P.O. Fedichev and G.V. Shlyapnikov, Phys. Rev. A {\bf 60}, R1779
(1999).



\bibitem{Stringari}
F. Zambelli and S. Stringari, Phys. Rev. Lett. {\bf 81}, 1754
(1998).

\bibitem{Dodd97} R. Dodd {\it et al.}, Phys. Rev. A {\bf 56}, 587 (1997).

\bibitem{Sinha97} S. Sinha, Phys. Rev. A {\bf 55}, 4325 (1997).

\bibitem{Svidzinsky98} A. Svidzinsky and A. Fetter, Phys. Rev. A. {\bf 58}, 3168 (1998).

\bibitem{Ohberg} P. Ohberg \emph{et al.}, Phys. Rev. A
\textbf{56}, R3346 (1997).

\bibitem{lz}
F. Chevy {\it et al.}, Phys. Rev. Lett. {\bf 85}, 2223 (2000).

\bibitem{Svidzinsky}
A.A. Svidzinsky and A.L. Fetter, Phys. Rev. A {\bf 62}, 063617
(2000).


\bibitem{Feder}
D.L. Feder \emph{et al.}, Phys. Rev. Lett. \textbf{86}, 564
(2001).


\bibitem{Rokshar}
D.A. Butts and D. S. Rokshar, Nature \textbf{397}, 327 (1999).

\bibitem{StringariModes}
S. Stringari, Phys. Rev. Lett. \textbf{77}, 2360-2363 (1996)

\bibitem{Bretin03}
V. Bretin, P. Rosenbusch, F. Chevy, G.V. Shlyapnikov, and J.
Dalibard, to be published.

\end{thebibliography}
\end{document}